\documentclass[useAMS,usenatbib,usegraphicx]{mn2e}
\def\etal{\textit{et al.}\ }

\def\pmi{$\pm$}

\title[Broad Band Optical Polarimetric Study of IC 1805]{Broad Band Optical Polarimetric Study of IC 1805}
\author[Biman J. Medhi, Maheswar G., Brijesh K., J.C. Pandey, T.S. Kumar and Ram Sagar]
{Biman J. Medhi$^{1}$\thanks{E-mail: biman@aries.ernet.in},Maheswar G.$^{1}$,Brijesh K.$^{1}$,J.C. Pandey$^{2}$,
T.S. Kumar$^{1}$,Ram Sagar$^{1}$ \\
$^{1}$Aryabhatta Research Institute of Observational Sciences, Manora Peak, 
Nainital - 263 129, India\\
$^{2}$Tata Institute of Fundamental Research, Homi Bhabha Road,
Mumbai - 400 005 , India} 
\begin{document}

\date{}

\pubyear{2007}

\maketitle

\label{firstpage}

\begin{abstract}

We present the $BVR$ broad  band polarimetric observations of 51 stars
belonging  to  the  young  open  cluster  IC  1805.   Along  with  the
photometric data  from the literature  we have modeled  and subtracted
the  foreground  dust   contribution  from  the  maximum  polarization
($P_{max}$)  and colour  excess  ($E_{B-V}$). The  mean  value of  the
$P_{max}$ for ${\it intracluster}$ medium and the foreground are found
to  be $5.008  \pm \  0.005 \  \%  $ and  $4.865 \pm  \ 0.022  \ \%  $
respectively.  Moreover,  the mean value of the  wavelength of maximum
polarization  ($\lambda_{max}$)  for  ${\it intracluster}$  medium  is
$0.541 \pm \  0.003 \ \mu m  $, which is quite similar  as the general
interstellar  medium (ISM).  The resulting  ${\it  intracluster}$ dust
component  is  found to  have  negligible  polarization efficiency  as
compared to interstellar dust. Some of the observed stars in IC 1805 have shown
the indication of intrinsic polarization in their measurements.

\end{abstract}

\begin{keywords}
ISM: dust, extinction, polarization - open clusters: individual (IC 1805)
\end{keywords}

\section{Introduction}
Polarization of starlight is  one  among  a  number of  properties  manifested  by
interstellar dust grains.   Wavelength dependence of polarization, for
instance,  can give  information on  the size  distribution  of grains
towards different  Galactic directions. Polarization is  thought to be
caused  by  the same  dust  grain  responsible  for the  reddening  of
starlight. According  to Davis and Greenstein  mechanism (Davis \etal,
1951),  the  polarization of  starlight  is  caused  by the  selective
extinction due to the elongated  dust grains aligned in space possibly
due to magnetic field.

In order  to investigate the  distribution and characteristic  of dust
grains,  we  have  selected  the  young open  cluster  IC  1805.   The
reddening law is anomalous in this direction and the extinction is non
uniform  over the cluster  region (Sagar  1987; Borgman  1961; Johnson
1968; Ishida 1969; Turner 1976).  The polarimetric components produced
by the Galactic dust located in  front of the cluster can be found out
from  maximum  polarization  $P_{max}$  and  colour  excess  $E_{B-V}$
(Marraco \etal 1993).   By removing the effect of  dust located on the
line of  sight from the data,  one can study  the component associated
with the internal extinction of  the cluster.  The young and rich open
cluster   IC  1805   $(R.A.(J2000):02^{h}\  32^{m}   \   42^{s},  Dec\
(J2000):+61^{d}\  27^{m}\   00^{s})$  is   the  core  of   the  CasOB6
association. It is located in the Perseus spiral arm, radially outward
from the  local spiral arm.  IC 1805  is situated in a  HII region and
embedded in W4 molecular cloud.  The colour excess $E_{(B-V)}$ for the
cluster  members  varies  from  0.52   to  1.3  mag  (Joshi  \&  Sagar
1983). From  the near-IR  study of IC  1805, Sagar \etal  (1990) found
that the distribution of dust and  patchy ionized gas appear to be the
cause of non uniform extinction across the cluster face.  The relative
proper-motion cluster  membership study was done  by Vasilevskis \etal
(1965) in a  field of  about $0.66$ square  degree, centered on  the O
type  giant  VSA 148  for  350  stars with  a  mean  error  of $\pm  \
0.16^{\prime\prime}$/century. Sanders  (1972) revised these membership
probabilities by  applying the maximum-likelihood  method.  The recent
estimation of  cluster distance  is $\simeq 2.4$  kpc (Joshi  \& Sagar
1983), though the earlier estimate varies from 1.6 to 2.5 kpc.

The previous polarimetric study on  IC 1805 was carried out by Guetter
\etal (1989),  for 24 member  stars brighter than visual  magnitude of
$\simeq$ 12 mag.  In order  to study details about the grain alignment
and particle size distribution we  have observed 32 member stars (nine
star common with Guetter \etal 1989) brighter than visual magnitude of
$\simeq$ 15.5 mag in IC  1805.  The next section describes the details
of  instrumentation, observation  and  data reduction  while the  main
results are given and discussed in the remaining part of the paper.

\section{Instrumentation and observation}

The ARIES Imaging  Polarimeter (AIMPOL) consists of a  half wave plate
(HWP) modulator  and a beam-splitting Wollaston  prism analyzer placed
in  the telescope  beam  path to  produce  ordinary and  extraordinary
images  in slightly  different  directions separated  by  about $27  \
pixel$. 
\begin{figure*}
\centering
\includegraphics[scale = .37, trim = 00 00 00 00, clip]{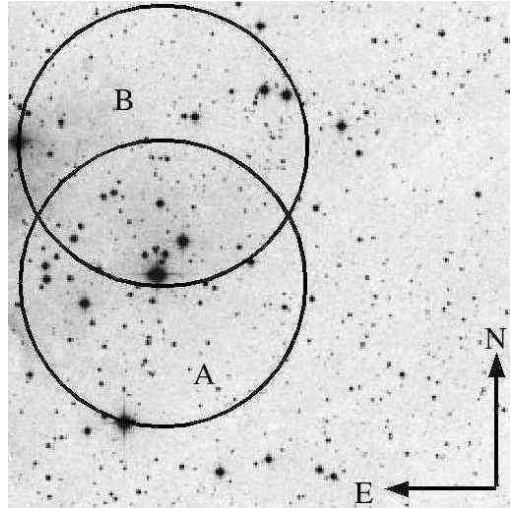}
\vskip.5cm
\caption{The $11.00^{\prime}\times11.00^{\prime}$ B-band image of the fields A \& B of IC 1805, reproduced from Digitized Sky Survey.}
\label{fig1}
\end{figure*}

\begin{table*}
\centering
\begin{minipage}{110mm}
\caption{Observed polarized standard stars}
\begin{tabular}{llllll}
\hline
Star name & Filter  &$P$\pmi$\epsilon_{P}$$(\%)$ &  $\theta$\pmi $\epsilon_{\theta}$$(^\circ)$ &  $P$\pmi$\epsilon_{P}$$(\%)$ &  
$\theta$\pmi$\epsilon_{\theta}$$(^\circ)$ \\
\hline
 & & \multicolumn{2}{c}{Published data}&\multicolumn{2}{c}{This paper}\\
\hline
Hiltner-960       &B & 5.72\pmi0.06 & 55.06\pmi 0.31  & 5.60\pmi 0.21&  54.62 \pmi 1.08 \\
                  &V & 5.66\pmi0.02 & 54.79\pmi 0.11  & 5.71\pmi 0.15&  53.32 \pmi 0.05 \\ 
                  &R & 5.21\pmi0.03 & 54.54\pmi 0.16  & 5.19\pmi 0.06&  54.81 \pmi 0.34 \\ 
HD204827          &B & 5.65\pmi0.02 & 58.20\pmi 0.11  & 5.75\pmi 0.09&  58.63 \pmi 0.45 \\ 
                  &V & 5.32\pmi0.02 & 58.73\pmi 0.08  & 5.36\pmi 0.03&  60.19 \pmi 0.17 \\
                  &R & 4.89\pmi0.03 & 59.10\pmi 0.17  & 4.90\pmi 0.23&  58.82 \pmi 1.14 \\
BD+64$^{\circ}$106&B & 5.51\pmi0.09 & 97.15\pmi 0.47  & 5.44\pmi 0.10&  99.42 \pmi 0.51 \\
                  &V & 5.69\pmi0.04 & 96.63\pmi 0.18  & 5.49\pmi 0.13&  97.06 \pmi 0.17 \\
                  &R & 5.15\pmi0.10 & 96.74\pmi 0.54  & 5.21\pmi 0.02&  97.38 \pmi 0.11 \\
HD19820           &B & 4.70\pmi0.04 & 115.70\pmi 0.22 & 4.84\pmi 0.23& 113.42 \pmi 0.11 \\
                  &V & 4.79\pmi0.03 & 114.93\pmi 0.17 & 4.92\pmi 0.11& 114.50 \pmi 0.11 \\
                  &R & 4.53\pmi0.03 & 114.46\pmi 0.17 & 4.73\pmi 0.14& 113.82 \pmi 0.28 \\
\hline										    					  
\end{tabular}
\end{minipage}
\end{table*}

A focal  reducer (85mm,  f/1.8) is  placed  between Wollaston
prism  and CCD  camera. The  detail descriptions  about the  AIMPOL are
available in Rautela \etal (2004) (also see Ramaprakash \etal 1998).

By definition,  the ratio R($\alpha$) is given by,
\begin{equation}
R(\alpha) = {I_{\circ}/I_{e} - 1 \over I_{\circ}/I_{e} + 1} = P\ cos(2\theta - 4\alpha)
\end{equation}
which  is  the difference  between  the  intensities  of the  ordinary
($I_{\circ}$) and extraordinary ($I_{e}$) beams to their sum, P is the
fraction of  the total light  in the linearly polarized  condition and
$\theta$  is the  position  angle  of plane  of  polarization.  It  is
denoted by normalized Stokes' parameter {\it q}\ (=Q/I), when the half
wave plate's  fast axis  is aligned to  the reference axis  ($\alpha =
0^\circ$).   Similarly,  the normalized  Stokes'  parameter ${\it  u}\
(=U/I), {\it q_{1}}\ (=Q_{1}/I), {\it u_{1}}\ (=U_{1}/I)$ are also the
ratios  R($\alpha$), when  the half  wave plate  are  at $22.5^\circ$,
$45^\circ$ and $67.5^\circ$ respectively. In principle, P and $\theta$
can be  determined by  using only two  Stokes' parameters {\it  q} and
{\it u}.  In reality, the  situation is not  so simple because  of two
reasons  (i) the  responsivity of  the  system to  the two  orthogonal
polarization components may  not be same and (ii)  the responsivity of
the CCD  is a function  of position on  its surface.  So,  the signals
which are  actually measured  in the two  images ($I_{\circ}^{\prime}\
and  \  I_{e}^{\prime}$) may  differ  from  the  observed one  by  the
following formula (Ramaprakash 1998)

\begin{equation}
{I_{\circ}^{\prime}(\alpha) \over I_{e}^{\prime}(\alpha)} = {I_{\circ}(\alpha) \over I_{e}(\alpha)} \times  {F_{\circ} \over F_{e}}
\end{equation}
and the Eq.(1) can be rewrite as

\begin{equation}
R(\alpha) = {(I_{\circ}/I_{e}\times F_{\circ}/{F_{e}})  - 1 \over (I_{\circ}/I_{e}\times F_{\circ}/{F_{e}}) + 1} = P\ cos(2\theta - 4\alpha)
\end{equation}
where $F_{\circ}$ and $F_{e}$ represent the effects mentioned above and the ratio is given by,

\begin{equation}
{ F_{\circ} \over {F_{e}}} = \left[{I_{\circ}^{\prime} ({0}^{\circ}) \over {I_{e}^{\prime} ({45}^{\circ})}}
\times {I_{\circ}^{\prime} ({45}^{\circ}) \over {I_{e}^{\prime} ({0}^{\circ})}}
\times {I_{\circ}^{\prime} ({22.5}^{\circ}) \over {I_{e}^{\prime} ({67.5}^{\circ})}}
\times {I_{\circ}^{\prime} ({67.5}^{\circ}) \over {I_{e}^{\prime} ({22.5}^{\circ})}}\right]^{1/4}
\end{equation}
Substituted the  ratio in  Eq.(3) and fitting  the cosine curve  to the
four values of $R(\alpha)$, the values of P and $\theta$ could be obtained. 
The individual errors associated with the four values
of $R(\alpha)$ putting  as a weight while calculating  P, $\theta$ and
their respective errors.

\begin{table*}
\centering
\begin{minipage}{120mm}
\caption{Observed $BVR$ polarization values for different stars in IC 1805}
\begin{tabular}{llllllll}
\hline
\hline
Id &$M_{V}$ (mag) & $P_{B}$\pmi $\epsilon_{p}$ $(\%)$ & $\theta_{B}$ \pmi $\epsilon_{\theta}$ $(^{\circ})$ &  $P_{V}$ \pmi  
$\epsilon_{p}$ $(\%)$ & $\theta_{V} $\pmi $\epsilon_{\theta}$ $(^{\circ})$& $P_{R}$\pmi  $\epsilon_{p}$ $(\%)$ & $\theta_{R} $
\pmi $\epsilon_{\theta}$ $(^{\circ})$ \\
\hline										    							
$112^G$  & 09.91  &  6.42\pmi 0.09 &  121.3\pmi   0.4 &   6.80\pmi  0.09 &   121.1\pmi  0.1 &  6.40\pmi  0.14 &    120.7\pmi   0.6 \\ 	  
$118^G$  & 10.30  &  5.89\pmi 0.15 &  122.8\pmi   0.2 &   5.90\pmi  0.01 &   121.5\pmi  0.1 &  5.79\pmi  0.01 &    121.3\pmi   0.6 \\ 
121      & 11.59  &  6.20\pmi 0.18 &  126.8\pmi   0.7 &   6.35\pmi  0.09 &   126.7\pmi  0.3 &  5.96\pmi  0.09 &    126.4\pmi   0.4 \\
122      & 13.73  &  4.99\pmi 0.61 &  120.6\pmi   3.6 &   5.10\pmi  0.51 &   118.5\pmi  0.1 &  5.04\pmi  0.43 &    118.6\pmi   2.4 \\ 
123      & 13.88  &  2.90\pmi 0.14 &  121.6\pmi   1.4 &   3.84\pmi  0.18 &   120.2\pmi  1.3 &  3.05\pmi  0.15 &    121.6\pmi   0.3 \\
128      & 13.40  &  3.41\pmi 0.12 &  108.7\pmi   0.8 &   3.92\pmi  0.17 &   120.0\pmi  1.0 &  3.80\pmi  0.09 &    118.8\pmi   0.6 \\
129      & 14.06  &  5.07\pmi 0.18 &  123.4\pmi   1.3 &   5.78\pmi  0.14 &   126.2\pmi  1.2 &  5.21\pmi  0.18 &    124.5\pmi   1.4 \\
$130^*$  & 13.26  &  4.75\pmi 0.30 &  124.0\pmi   0.5 &   5.53\pmi  0.03 &   124.5\pmi  1.0 &  5.15\pmi  0.04 &    126.5\pmi   0.1 \\
$133^*$  & 13.56  &  3.28\pmi 0.38 &  118.3\pmi   3.3 &   3.64\pmi  0.19 &   125.9\pmi  1.5 &  3.30\pmi  0.48 &    123.4\pmi   4.1 \\
$136^G$  & 11.04  &  5.39\pmi 0.10 &  121.9\pmi   0.5 &   5.41\pmi  0.09 &   119.9\pmi  0.4 &  5.38\pmi  0.04 &    120.9\pmi   0.2 \\
$138^G$  & 09.58  &  5.80\pmi 0.08 &  122.9\pmi   0.4 &   6.31\pmi  0.08 &   119.0\pmi  0.3 &  5.40\pmi  0.36 &    120.0\pmi   2.3 \\
139      & 13.10  &  5.12\pmi 0.04 &  123.1\pmi   0.2 &   5.30\pmi  0.17 &   119.3\pmi  0.8 &  5.10\pmi  0.07 &    121.0\pmi   0.4 \\ 
143      & 11.40  &  4.68\pmi 0.04 &  125.4\pmi   0.2 &   5.25\pmi  0.05 &   122.5\pmi  0.2 &  4.80\pmi  0.07 &    122.6\pmi   0.4 \\
$146^*$  & 13.53  &  5.44\pmi 0.56 &  124.7\pmi   2.9 &   5.72\pmi  0.34 &   117.3\pmi  1.7 &  4.97\pmi  0.07 &    119.4\pmi   0.4 \\
147      & 13.34  &  4.23\pmi 0.30 &  122.6\pmi   3.2 &   4.86\pmi  0.15 &   122.1\pmi  0.8 &  4.71\pmi  0.17 &    122.2\pmi   0.9 \\
$149^G$  & 11.24  &  4.37\pmi 0.09 &  122.4\pmi   0.5 &   4.93\pmi  0.09 &   119.6\pmi  0.5 &  4.56\pmi  0.05 &    118.3\pmi   0.3 \\
154      & 14.07  &  4.19\pmi 0.59 &  114.7\pmi   3.8 &   4.27\pmi  0.23 &   115.6\pmi  1.5 &  4.07\pmi  0.25 &    123.8\pmi   1.8 \\
$155^*$  & 14.39  &  4.95\pmi 0.66 &  111.5\pmi   3.8 &   4.52\pmi  0.17 &   116.6\pmi  1.0 &  4.35\pmi  0.09 &    111.3\pmi   0.6 \\
$156^G$  & 12.03  &  4.19\pmi 0.05 &  122.2\pmi   0.3 &   4.28\pmi  0.04 &   120.6\pmi  0.2 &  3.55\pmi  0.17 &    122.6\pmi   1.3 \\
157      & 13.48  &  5.17\pmi 0.29 &  121.4\pmi   1.2 &   5.67\pmi  0.30 &   119.8\pmi  1.5 &  5.54\pmi  0.05 &    120.8\pmi   0.2 \\
158      & 12.73  &  3.98\pmi 0.11 &  122.9\pmi   4.4 &   4.50\pmi  0.10 &   126.1\pmi  0.6 &  4.27\pmi  0.02 &    127.5\pmi   0.1 \\
$160^G$  & 08.11  &  5.58\pmi 0.19 &  117.4\pmi   0.9 &   5.91\pmi  0.46 &   117.5\pmi  2.2 &  5.30\pmi  0.06 &    117.4\pmi   0.3 \\
162      & 12.55  &  5.63\pmi 0.46 &  121.9\pmi   2.3 &   5.77\pmi  0.08 &   118.3\pmi  0.4 &  5.57\pmi  0.22 &    119.5\pmi   1.2 \\
165      & 12.77  &  5.23\pmi 0.14 &  122.4\pmi   0.7 &   5.68\pmi  0.16 &   119.0\pmi  0.8 &  5.18\pmi  0.04 &    119.3\pmi   0.2 \\
166      & 11.97  &  4.74\pmi 0.31 &  119.6\pmi   1.3 &   4.99\pmi  0.29 &   119.2\pmi  0.2 &  4.49\pmi  0.14 &    120.5\pmi   0.2 \\
168      & 13.72  &  6.06\pmi 0.82 &  122.0\pmi   0.8 &   5.29\pmi  0.34 &   117.1\pmi  1.8 &  5.16\pmi  0.10 &    119.3\pmi   0.5 \\
170      & 10.07  &  3.52\pmi 0.09 &  121.2\pmi   0.1 &   3.75\pmi  0.07 &   118.3\pmi  0.5 &  3.44\pmi  0.26 &    118.8\pmi   2.1 \\
171      & 13.13  &  5.16\pmi 0.13 &  124.0\pmi   0.6 &   5.24\pmi  0.14 &   115.4\pmi  2.0 &  5.13\pmi  0.05 &    119.5\pmi   0.3 \\
$173^*$  & 13.81  &  2.67\pmi 0.12 &  112.3\pmi   1.2 &   2.62\pmi  0.09 &   116.0\pmi  1.0 &  2.59\pmi  0.09 &    118.8\pmi   0.1 \\
174      & 11.63  &  5.29\pmi 0.17 &  119.5\pmi   0.8 &   5.27\pmi  0.01 &   118.3\pmi  0.1 &  5.06\pmi  0.02 &    117.9\pmi   0.1 \\
175      & 13.10  &  4.68\pmi 0.44 &  114.4\pmi   2.5 &   4.96\pmi  0.24 &   115.1\pmi  1.3 &  4.48\pmi  0.14 &    116.1\pmi   0.8 \\
182      & 13.38  &  5.61\pmi 0.05 &  117.0\pmi   0.1 &   5.94\pmi  0.09 &   122.9\pmi  1.0 &  5.53\pmi  0.02 &    121.1\pmi   0.2 \\
$183^G$  & 11.15  &  5.19\pmi 0.37 &  121.0\pmi   2.0 &   5.01\pmi  0.04 &   118.5\pmi  0.2 &  4.89\pmi  0.06 &    118.5\pmi   0.3 \\
$184^*$  & 13.64  &  4.20\pmi 0.28 &  120.4\pmi   1.6 &   3.82\pmi  0.11 &   120.0\pmi  0.7 &  3.81\pmi  0.16 &    124.6\pmi   1.2 \\
$185^*$  & 11.58  &  5.56\pmi 0.05 &  121.9\pmi   0.2 &   5.44\pmi  0.19 &   120.1\pmi  0.9 &  4.99\pmi  0.07 &    120.2\pmi   0.3 \\
188      & 12.68  &  4.90\pmi 0.12 &  122.8\pmi   0.8 &   5.38\pmi  0.09 &   121.2\pmi  0.4 &  4.76\pmi  0.18 &    118.1\pmi   0.4 \\
191      & 12.96  &  3.99\pmi 0.11 &  113.0\pmi   1.0 &   4.32\pmi  0.18 &   105.8\pmi  6.4 &  3.97\pmi  0.09 &    117.5\pmi   0.7 \\
$192^G$  & 08.43  &  3.38\pmi 0.16 &  123.0\pmi   1.0 &   3.52\pmi  0.11 &   115.0\pmi  0.8 &  3.31\pmi  0.18 &    118.4\pmi   1.6 \\
$354^*$  & 13.43  &  4.75\pmi 0.35 &  125.9\pmi   1.7 &   4.82\pmi  0.31 &   116.7\pmi  6.7 &  4.28\pmi  0.05 &    119.2\pmi   0.3 \\
$359^*$  & 14.55  &  2.82\pmi 0.89 &  103.8\pmi   9.0 &   4.10\pmi  0.30 &   117.4\pmi  1.6 &  3.76\pmi  0.15 &    122.6\pmi   1.1 \\
$360^*$  & 14.52  &  3.14\pmi 0.41 &  121.3\pmi   3.6 &   3.10\pmi  0.09 &   123.2\pmi  0.8 &  3.31\pmi  0.45 &    123.2\pmi   3.8 \\
$362^*$  & 14.48  &  4.66\pmi 0.25 &  121.4\pmi   2.8 &   5.46\pmi  0.27 &   119.7\pmi  0.9 &  4.70\pmi  0.02 &    117.2\pmi   0.1 \\
$363^*$  & 14.61  &  4.77\pmi 0.41 &  101.8\pmi  14.8 &   5.77\pmi  0.27 &   118.0\pmi  1.3 &  4.60\pmi  0.30 &    119.7\pmi   2.1 \\
$365^*$  & 14.56  &  4.41\pmi 0.21 &  120.8\pmi   1.3 &   4.33\pmi  0.04 &   116.2\pmi  0.2 &  3.88\pmi  0.64 &    120.8\pmi   6.3 \\
$424^*$  & 14.41  &  4.61\pmi 0.68 &  122.7\pmi   4.3 &   4.57\pmi  0.02 &   124.9\pmi  0.1 &  4.46\pmi  0.02 &    116.4\pmi   0.1 \\
$426^*$  & 15.11  &  6.12\pmi 1.02 &  121.8\pmi   4.1 &   5.36\pmi  0.22 &   119.7\pmi  1.1 &  5.12\pmi  0.12 &    120.8\pmi   1.1 \\
$429^*$  & 14.92  &  5.19\pmi 0.94 &  120.8\pmi   5.2 &   4.82\pmi  0.14 &   113.6\pmi  1.0 &  4.96\pmi  0.16 &    121.4\pmi   2.5 \\
$470^*$  & 14.03  &  5.12\pmi 0.31 &  117.4\pmi   1.7 &   4.69\pmi  0.05 &   115.1\pmi  0.3 &  4.20\pmi  0.02 &    118.4\pmi   0.1 \\
$480^*$  & 15.47  &  7.84\pmi 1.52 &  116.9\pmi   8.9 &   6.56\pmi  0.93 &   123.0\pmi  4.0 &  6.78\pmi  0.45 &    120.7\pmi   1.9 \\
$541^*$  & 13.71  &  4.96\pmi 0.70 &  120.7\pmi   5.0 &   5.54\pmi  0.46 &   119.7\pmi  2.3 &  5.18\pmi  0.86 &    120.6\pmi   4.7 \\
$552^*$  & 14.47  &  7.39\pmi 0.05 &  110.1\pmi   0.2 &   7.28\pmi  0.13 &   113.6\pmi  4.4 &  6.86\pmi  0.58 &    118.9\pmi   2.4 \\
\hline
\end{tabular}
\end{minipage}
\begin{quote}{\hspace{2.5cm}{ $^*$ : Nonmember}} \\
{\hspace{2.6cm}{ $^G$ : Common with Guetter \etal 1989}} \\
\end{quote}
\end{table*}

The optical imaging polarimetry of the  two fields A and B (see Figure
1) in  IC  1805   was  carried  out  to  study   the  contribution  of
interstellar and ${\it intracluster}$ material on linear polarization.
The data were obtained on  12$^{th}$ and 13$^{th}$ October, 2006 using
the  TK  $1024\times  1024$   pixel$^2$  CCD  camera  mounted  on  the
Cassegrain  focus  of  the  104-cm Sampurnanand  telescope  of  ARIES,
Nainital  in  $B$, $V$  and  $R$  ($\lambda_{B_{eff}}$=0.440 $\mu  m$,
$\lambda_{V_{eff}}$=0.550  $\mu m$ and  $\lambda_{R_{eff}}$=0.660 $\mu
m$) photometric  bands. Each  pixel of the  CCD corresponds  to $1.73$
arcsec and the  field of view is $\sim 8$ arcmin  diameter on the sky.
The  FWHM of  the  stellar  image varies  from  $2 \  pixel$  to $3  \
pixel$. The read out noise and gain of the CCD are 7.0 $e^-$ and 11.98
$e^-$/ADU  respectively.  The fluxes  for all  of our  programme stars
were extracted  by aperture photometry  after the bias  subtraction in
the standard manner  using IRAF.  Instead of  robust flat fielding
technique we are  using Eq.[3] to make uniform  response, as mentioned
above.  

Standard stars  for null  polarization and for  the zero point  of the
polarization position angle were  taken from Schmidt \etal (1992). The
results for polarized standards are given in Table 1. From the results
we can conclude that the observed polarization and position angles are
matching with Schmidt \etal (1992) within the error limit. The average
value  of instrumental  polarization  is  found to  be  about $\sim  \
0.04\%$.

The AIMPOL  does not have  a grid placed  to avoid the  overlapping of
ordinary image with the extraordinary  image of an adjacent region $27
\  pixels$ away  from  it. In  the  care of  target  sources, we  have
selected (usually)  only those which  are well isolated.  But,  due to
the overlapping, sky at a region gets doubled.  However, we found the
sky    variation   at    different   location    to   be    not   very
significant. Therefore, the effect gets canceled when we consider both
ordinary  and extraordinary  images of  these target  sources  for the
analysis.

\begin{figure}
\centering
\includegraphics[scale = .4, trim = 00 15 15 15 , clip]{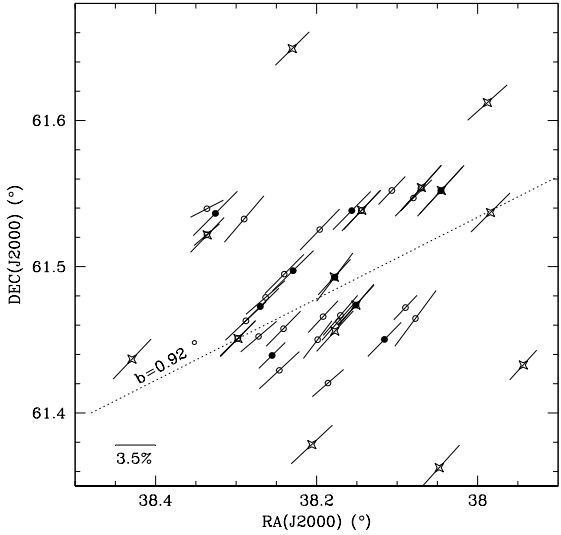}
\vskip.5cm
\caption{Polarization vectors and their orientations for the stars of IC 1805. The vectors are proportional to the magnitude of the polarization.
The scale is indicated in the figure. Filled and open circles indicate members and nonmembers of cluster IC 1805 observed by us and the starmark 
indicate the stars observed by Guetter \etal (1989). The dotted line is the galactic parallel $b= 0.92^{\circ}$}
\label{fig1}
\end{figure}
\begin{table}
\centering
\begin{minipage}{220mm}
\caption{ Polarization results for IC 1805 stars}
\begin{tabular}{lllll}
\hline
\hline
Id & \hspace{-3mm}{$P_{max}$ \pmi $\epsilon$} $(\%)$ &\ \  $\sigma_1$  & $\lambda_{max} $ \pmi $\epsilon$ ($\mu m$) & $E_{B-V}$\\
\hline
$112^\dag$& 6.76\pmi   0.04 &  0.66 & \ \ \ 0.54\pmi    0.01 &   0.82 \\	     
118      & 5.92\pmi   0.02 &  2.96 & \ \ \ 0.57\pmi    0.01 &   0.85 \\
121      & 6.38\pmi   0.01 &  0.05 & \ \ \ 0.52\pmi    0.01 &   0.88 \\
122      & 5.20\pmi   0.08 &  0.25 & \ \ \ 0.55\pmi    0.02 & \ ---  \\
123      & 3.32\pmi   0.33 &  3.40 & \ \ \ 0.58\pmi    0.13 &   0.79 \\ 
128      & 3.85\pmi   0.05 &  0.68 & \ \ \ 0.60\pmi    0.02 &   0.97 \\
$129^\dag$& 5.58\pmi   0.20 &  2.00 & \ \ \ 0.56\pmi    0.06 &   0.81 \\ 
$130^*$  & 5.54\pmi   0.09 &  2.21 & \ \ \ 0.51\pmi    0.02 & \ ---  \\ 
$133^*$  & 3.59\pmi   0.08 &  0.50 & \ \ \ 0.56\pmi    0.05 & \ ---  \\ 
136      & 5.58\pmi   0.10 &  2.23 & \ \ \ 0.55\pmi    0.02 &   0.89 \\
$138^\dag$& 6.26\pmi   0.18 &  2.06 & \ \ \ 0.56\pmi    0.04 &   0.79 \\
$139^\dag$& 5.35\pmi   0.01 &  0.31 & \ \ \ 0.54\pmi    0.01 &   0.76 \\ 
143      & 5.10\pmi   0.15 &  4.15 & \ \ \ 0.57\pmi    0.04 &   0.86 \\ 
$146^*$  & 5.69\pmi   0.27 &  0.70 & \ \ \ 0.47\pmi    0.03 & \ ---  \\ 
147      & 4.82\pmi   0.07 &  0.68 & \ \ \ 0.59\pmi    0.03 &   0.78 \\
$149^\dag$& 4.75\pmi   0.10 &  2.33 & \ \ \ 0.56\pmi    0.03 &   0.70 \\ 
154      & 4.30\pmi   0.03 &  0.13 & \ \ \ 0.53\pmi    0.01 &   0.96 \\ 
$155^*$  & 4.60\pmi   0.16 &  0.90 & \ \ \ 0.53\pmi    0.04 & \ ---  \\ 
156      & 4.29\pmi   0.08 &  2.41 & \ \ \ 0.50\pmi    0.03 &   0.88 \\ 
157      & 5.65\pmi   0.01 &  0.13 & \ \ \ 0.58\pmi    0.01 &   0.89 \\
158      & 4.38\pmi   0.06 &  1.45 & \ \ \ 0.57\pmi    0.02 &   1.24 \\ 
160      & 5.73\pmi   0.06 &  0.48 & \ \ \ 0.51\pmi    0.01 &   1.02 \\  
$162^\dag$& 5.78\pmi   0.13 &  0.36 & \ \ \ 0.54\pmi    0.02 &   0.67 \\ 
165      & 5.50\pmi   0.10 &  1.31 & \ \ \ 0.53\pmi    0.02 &   0.86 \\
166      & 4.90\pmi   0.09 &  0.51 & \ \ \ 0.50\pmi    0.02 &   0.75 \\
168      & 5.57\pmi   0.37 &  1.04 & \ \ \ 0.51\pmi    0.06 &   0.96 \\      	     	
$170\dag$& 3.73\pmi   0.04 &  0.63 & \ \ \ 0.55\pmi    0.01 &   0.35 \\      	     	
171      & 5.36\pmi   0.07 &  0.96 & \ \ \ 0.54\pmi    0.01 &   0.76 \\ 
$173^*$  & 2.70\pmi   0.06 &  1.11 & \ \ \ 0.53\pmi    0.04 & \ ---  \\      	     	
$174\dag$& 5.27\pmi   0.03 &  1.73 & \ \ \ 0.54\pmi    0.01 &   0.75 \\      	     	
175      & 4.92\pmi   0.12 &  0.53 & \ \ \ 0.50\pmi    0.02 &   0.85 \\      	     	
182      & 5.85\pmi   0.03 &  1.15 & \ \ \ 0.53\pmi    0.01 &   0.81 \\ 
183      & 5.02\pmi   0.05 &  1.40 & \ \ \ 0.56\pmi    0.03 &   0.88 \\
$184^*$  & 3.96\pmi   0.22 &  1.82 & \ \ \ 0.51\pmi    0.08 & \ ---  \\ 
$185^*$  & 5.61\pmi   0.01 &  0.25 & \ \ \ 0.48\pmi    0.01 & \ ---  \\ 
$188\dag$& 5.26\pmi   0.15 &  2.19 & \ \ \ 0.54\pmi    0.05 &   0.78 \\ 
191      & 4.19\pmi   0.05 &  0.77 & \ \ \ 0.54\pmi    0.02 &   0.88 \\
192      & 3.52\pmi   0.01 &  0.11 & \ \ \ 0.53\pmi    0.01 &   0.76 \\
$354^*$  & 4.84\pmi   0.09 &  0.34 & \ \ \ 0.48\pmi    0.01 & \ ---  \\
$359^*$  & 3.90\pmi   0.26 &  1.15 & \ \ \ 0.56\pmi    0.11 & \ ---  \\
$360^*$  & 3.12\pmi   0.08 &  0.86 & \ \ \ 0.55\pmi    0.10 & \ ---  \\
$362^*$  & 5.01\pmi   0.26 &  1.88 & \ \ \ 0.52\pmi    0.05 & \ ---  \\
$363^*$  & 5.36\pmi   0.53 &  2.33 & \ \ \ 0.50\pmi    0.12 & \ ---  \\ 
$365^*$  & 4.44\pmi   0.01 &  0.06 & \ \ \ 0.47\pmi    0.01 & \ ---  \\
$424^*$  & 4.58\pmi   0.01 &  0.54 & \ \ \ 0.57\pmi    0.01 & \ ---  \\ 
$426^*$  & 5.47\pmi   0.24 &  0.89 & \ \ \ 0.52\pmi    0.05 & \ ---  \\
$429^*$  & 4.95\pmi   0.12 &  1.01 & \ \ \ 0.62\pmi    0.06 & \ ---  \\ 
$470^*$  & 4.88\pmi   0.08 &  0.85 & \ \ \ 0.46\pmi    0.01 & \ ---  \\ 
$480^*$  & 7.24\pmi   0.70 &  0.87 & \ \ \ 0.51\pmi    0.09 & \ ---  \\
$541^*$  & 5.46\pmi   0.11 &  0.29 & \ \ \ 0.57\pmi    0.04 & \ ---  \\
$552^*$  & 7.45\pmi   0.02 &  0.41 & \ \ \ 0.48\pmi    0.01 & \ ---  \\     	
\hline
\end{tabular}
\end{minipage}
\begin{quote}{\hspace{.02cm}{$^*$ : Nonmember}} \\
{\hspace{.2cm}{$^\dag$ : Frontside\ stars }}
\end{quote}
\end{table}

\section{Results}

Table 2 lists, for the 51  observed stars in the direction of IC 1805,
the percentage  of polarization,  the position angle  of the  plane of
polarization in the equatorial  coordinate system and their respective
errors for each filters. Membership probabilities, star identification
numbers (Id) for all observed  stars are taken from Sanders (1972) and
the  visual  magnitude  $M_{V}$  taken  from  the  work  of  Joshi  \&
Sagar(1983)  and  Massey  \etal  (1995).  The  stars  with  membership
probability  greater than  $50\%$ are  consider as  the members  of IC
1805.  Our results for the common  stars are in agreement with Guetter
\etal  (1989) within  the  error  limit except  $P_{B}$.   There is  a
systematic increments of $\sim 0.45  \ \%$ in $P_{B}$ measurements for
our values from those of Guetter \etal (1989).

The polarization map  for all the stars of IC 1805  observed by us and
Guetter  \etal  (1989)  is shown  in  Figure  2.   The length  of  the
polarization vectors  are proportional  to the degree  of polarization
($P_{V}$) in  $ \% $  (scale is given  in the polarization  map).  The
polarization vectors make an angle ($\theta_{V}$) with the North-South
axis (reference  axis), which  is given  in column 6  of Table  2. The
filled and open  circles indicate the cluster member  and nonmember of
IC 1805 observed by us and the starmark indicate the stars observed by
Guetter \etal (1989).  From the  polarization map we can conclude that
the alignment of the polarization vectors over the whole observed area
are close to being parallel to  each other and there is no significant
difference  between the  alignment  of  the star  observed  by us  and
Guetter \etal (1989).  The alignment  of member and nonmember stars of
cluster  IC  1805  are  also  closely parallel  to  each  other.   The
orientation of the  polarization vectors for the observed  stars of IC
1805 indicate that the magnetic  field in the direction of the cluster
follow  a  general  trend   of  the  polarization  directions  in  the
region. The  polarization in the  cluster has an average  direction of
$\overline \theta_{V}$  = 119.96  $\pm$ 0.05 in  equatorial coordinate
system.  The dotted line  superimposed on  the Figure  2 is  the galactic
parallel   $b=0.92^{\circ}$   showing  a   close   alignment  of   the
polarization vectors with the projection of the Galactic plane.

\begin{figure}
\centering
\includegraphics[scale = .85, trim = 5 10 250 10, clip]{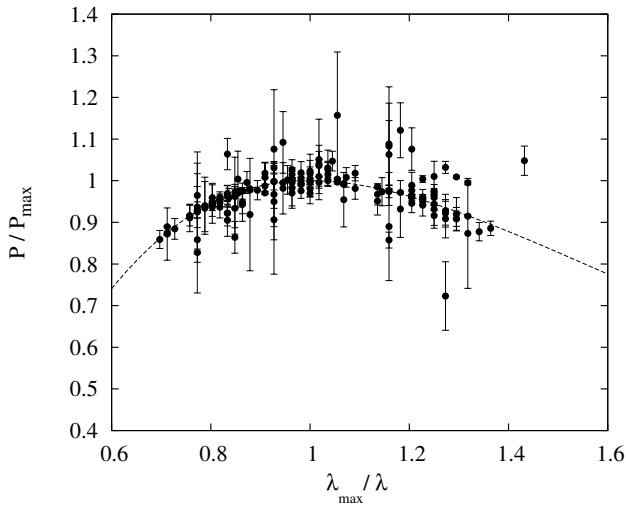}
\vskip.5cm
\caption{Plot of normalized polarization-wavelength dependence for the observed stars
in IC 1805 determined from data listed in Tables 2 and 3. The error bars represent the uncertainties
for the measurements of $P/P_{max}$ and the solid curve denotes the Serkowski polarization relation
for general interstellar medium (ISM) [Eq. (5)].}
\label{fig1}
\end{figure}
\begin{figure}
\centering
\includegraphics[scale = .4, trim = 00 00 00 00, clip]{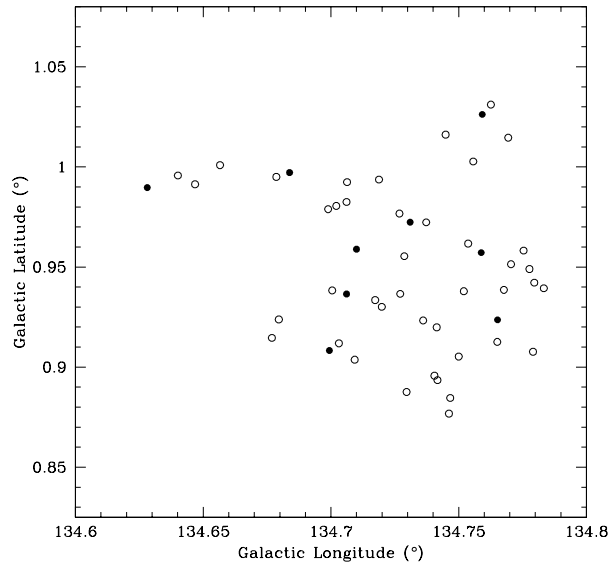}
\caption{Galactic distribution for all observed stars of IC 1805. Filled circles indicate the {\it frontside} stars. }
\label{fig1}
\end{figure}

\begin{figure}
\centering  \includegraphics[scale   =  .4,  trim  =  00   00  00  00, clip]{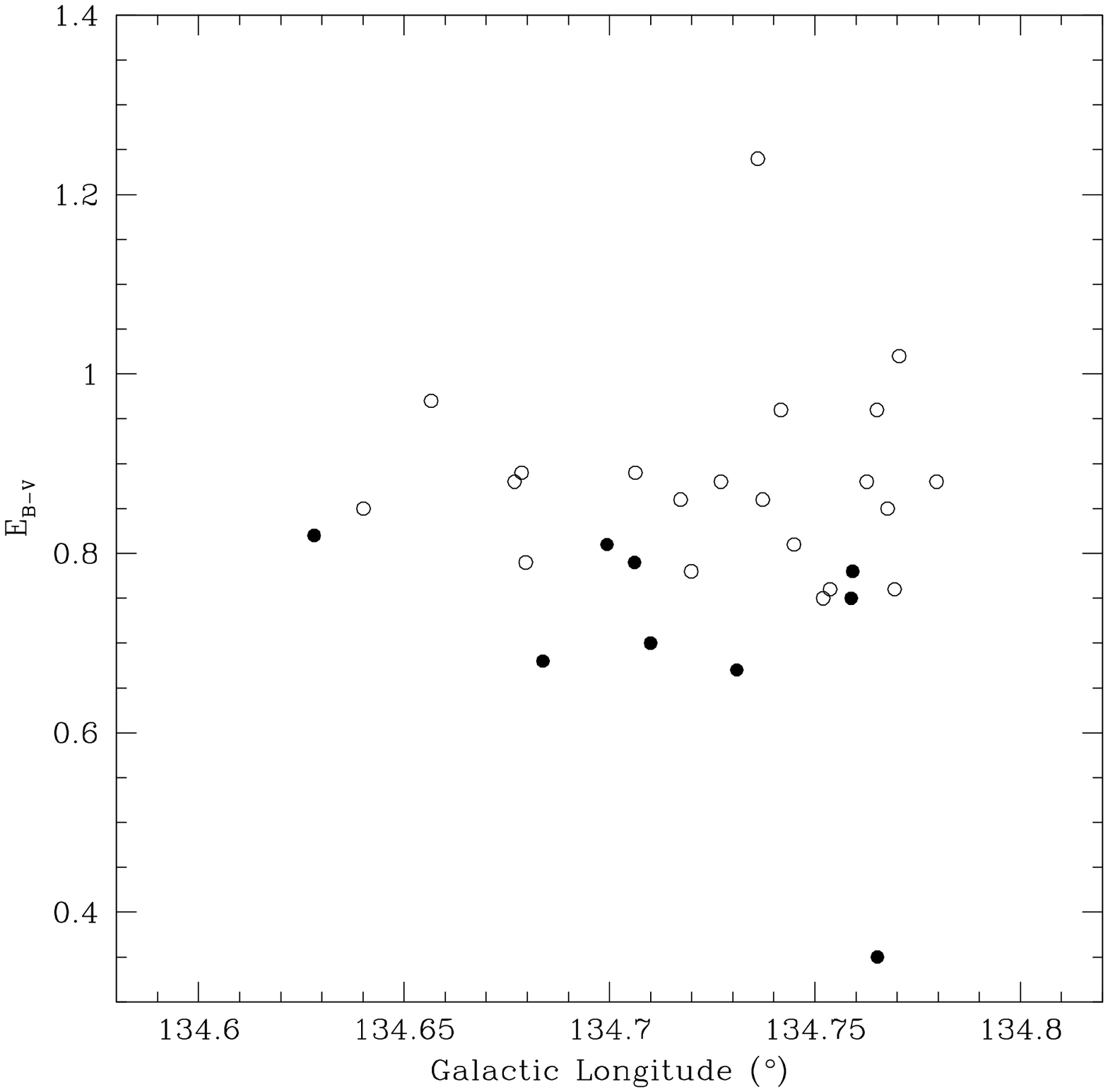}
\includegraphics[scale  =  .4,   trim  =  00  00  00  00, clip]{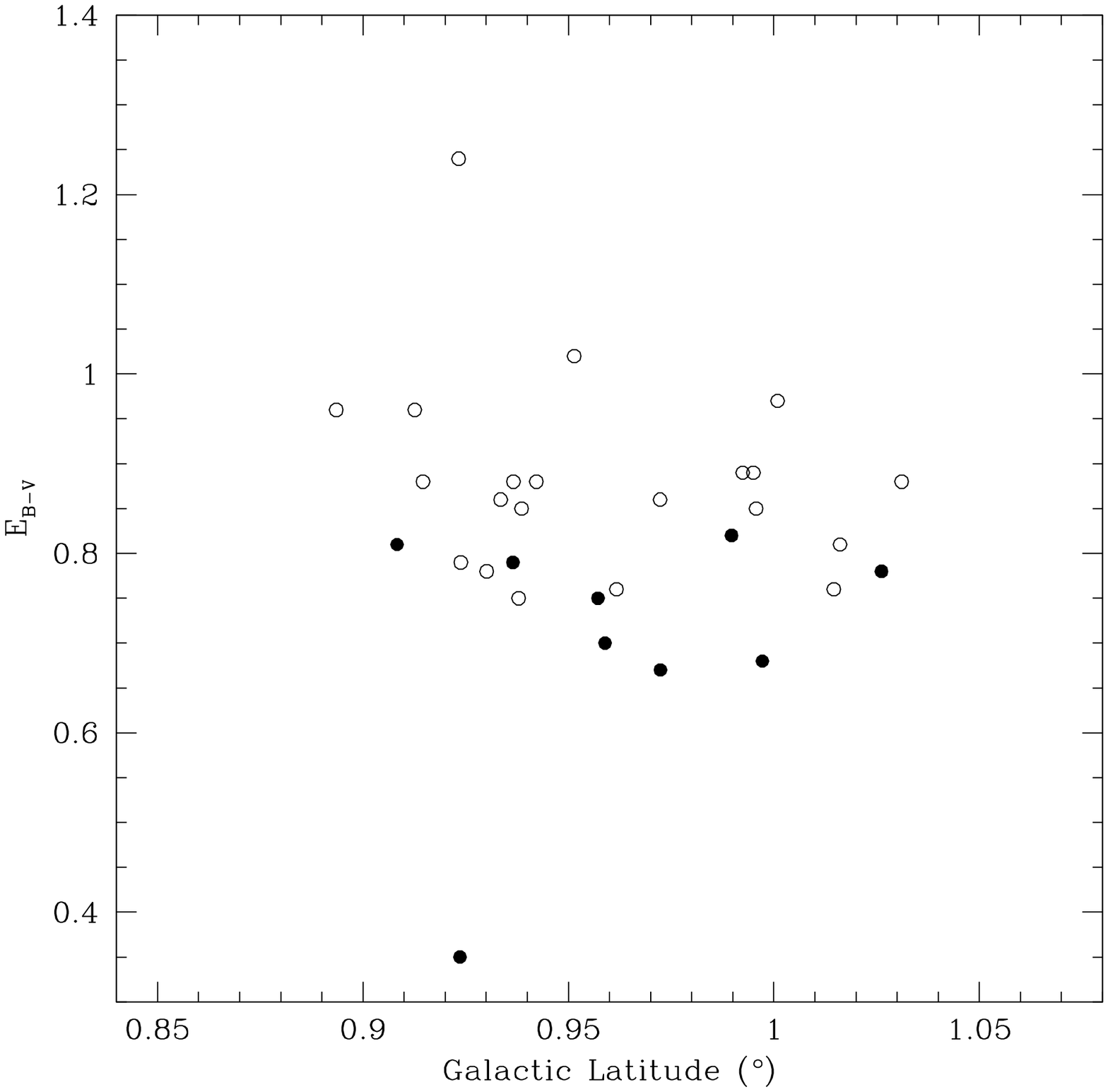}
\caption{Top (5a): Colour excess $E_{B-V}$ plotted as a function of galactic longitude ({\it l}). Bottom (5b): Colour excess $E_{B-V}$ plotted as a function of galactic latitude ({\it b}). Excesses are taken from Joshi \& Sagar (1983). Open circles indicate the observed member stars and filled circles indicate the {\it frontside} stars. }
\label{fig1}
\end{figure}

\section{Dust properties} 

The  wavelength   $(\lambda_{max})$  at  which   maximum  polarization
$(P_{max})$  occurs  is  a  function  of the  optical  properties  and
characteristic  of  particle size  distribution  of  the aligned  dust
grains  (Wilking \etal  1980;  McMillan 1978).  Moreover,  it is  also
related  to the  interstellar  extinction law  (Serkowski \etal  1975;
Whittet  \etal  1978; Coyne  \etal  1979;  Clayton  \etal 1988).   The
maximum wavelength $\lambda_{max}$ and polarization $P_{max}$ have been
calculated  by   fitting  the  observed  polarization   in  the  $BVR$
band-passes to the standard Serkowski's polarization law,

\begin{equation}
P_{\lambda}/ P_{max} =\ exp \left[-\ k \ ln^{2} \ (\lambda_{max}/\lambda) \right]
\end{equation}
and  adopting the  parameter $k  = 1.15  $ (Serkowski  1973).   If the
polarization is well represented by the Serkowski relation, $\sigma_1$
(the unit weight  error of the fit) should not be  higher than 1.6 due
to the  weighting scheme;  a higher value  could be indicative  of the
presence of intrinsic polarization.   In fitting the degree of freedom
is  adopted as  one.  Though there  are  only three  data points,  the
wavelength  convey ranges  from 0.44  to  0.66 $\mu  m $  and all  the
$\lambda_{max}$ found to fall within this range.

Table 3  represent $P_{max}$,  $\lambda_{max}$ and $\sigma_1$  for all
the 51 stars with their respective errors. The colour excess $E_{B-V}$
for  different observed  stars are  taken from  the work  of  Joshi \&
Sagar(1983). Of the 51 stars observed, 14 of them have the unit weight
error of  the fit above the  limiting value of 1.6  (member stars 174,
156, 143, 149, 138, 129, 123,  188, 136, 118 and non-member stars 363,
362, 130,  184).  As mentioned  above, the higher value  of $\sigma_1$
gives a clue about the presence of intrinsic polarization in the light
from the star.  So, in principle,  in case of IC 1805, interstellar dust alone does
not appear to be responsible  for the observed polarization.  There is
also a  indication of intrinsic polarization in  measurements for some
of the observed stars.

The  observed  normalized  polarization-wavelength dependence  of  the
stars in  IC 1805 is shown in  Figure 3.  The solid  curve denotes the
Serkowski polarization  relation for the  general interstellar medium.
In  this  plot,  there  exist  a good  agreement  between  theory  and
observations, which  indicates that the observed  polarization for the
stars of IC 1805 is mainly due to Davis-Greenstein mechanism (Davis \&
Greenstein 1951) that operates in the general interstellar medium.

The  foreground reddening toward  IC 1805  is $\sim$  0.8 mag  and the
reddening law  is anomalous in  this direction (Borgman  1961; Jhonson
1968;  Ishida 1969; Turner  1976).  In  the direction  of IC  1805 the
interstellar reddening  increases quite linearly  up to 1 kpc  with an
average  absorption of  about  0.6 mag/kpc.  This  indicates that  the
distribution  of interstellar  material  in the  local  arm is  rather
uniform. To remove the effect of interstellar dust located in front of
IC 1805 or to determine the effect of ${\it intracluster}$ dust on the
members of  IC 1805, we follow  the method described  by Marraco \etal
(1993). In order to  do the same, we have selected a  group of 9 stars
among the members  of IC 1805, covering the  whole observed region and
which  seems to  be  least  affected by  the  reddening.  These  stars
hereafter referred  to as  {\it frontside} stars,  they are indicated by
$^\dag$ in Table  3.  The {\it frontside} stars will  be used to model
the contribution of foreground  excess, polarization and finally which
will be removed from the  remaining member stars to determine the {\it
intracluster} parameters respectively.
\begin{table}
\centering
\begin{minipage}{280mm}
\caption{BVR polarization results }
\begin{tabular}{lllllll}
\hline
\hline
Id & $P_{max}$ & ${P_{max}}$ & ${P_{max}}$ & $E_{B-V}$& $E_{B-V}$& $E_{B-V}$\\
\hline
 &\multicolumn{1}{c}{o}&\multicolumn{1}{c}{f} & \multicolumn{1}{c}{i} & \multicolumn{1}{c}{o}&\multicolumn{1}{c}{f} & \multicolumn{1}{c}{i} \\
\hline
118&  5.92&   4.82&  1.10&   0.85& 0.72 & 0.13  \\
121&  6.38&   4.38&  2.00&   0.88& 0.72 & 0.16	\\
122&  5.20&   4.77&  0.43& \ --- & 0.72 & \ --- \\
123&  3.32&   1.78&  1.54&   0.79& 0.72 & 0.07	\\
128&  3.85&   2.84&  1.01&   0.97& 0.72 & 0.25	\\
136&  5.58&   4.86&  0.72&   0.89& 0.72 & 0.17	\\
143&  5.09&   4.60&  0.49&   0.86& 0.72 & 0.14	\\
147&  4.82&   4.46&  0.36&   0.78& 0.72 & 0.06	\\
154&  4.29&   3.39&  0.90&   0.96& 0.72 & 0.24	\\
156&  4.29&   3.71&  0.58&   0.88& 0.72 & 0.16	\\
157&  5.65&   4.86&  0.79&   0.89& 0.73 & 0.16	\\
158&  4.38&   3.29&  1.09&   1.24& 0.72 & 0.52	\\
160&  5.73&   4.75&  0.98&   1.02& 0.73 & 0.29 \\
165&  5.53&   4.84&  0.69&   0.86& 0.73 & 0.13 \\
166&  4.90&   4.76&  0.14&   0.75& 0.73 & 0.02 \\
168&  5.57&   4.69&  0.88&   0.96& 0.72 & 0.24 \\
171&  5.37&   4.41&  0.97&   0.76& 0.73 & 0.03	\\
175&  4.92&   4.08&  0.84&   0.85& 0.73 & 0.12 \\
182&  5.85&   4.72&  1.13&   0.81& 0.73 & 0.08	\\
183&  5.03&   4.72&  0.31&   0.88& 0.73 & 0.15 \\ 
191&  4.24&   2.31&  1.93&   0.88& 0.73 & 0.15	\\
192&  3.51&   1.98&  1.53&   0.76& 0.73 & 0.03  \\
\hline
\end{tabular}
\end{minipage}
\begin{quote}{
\hspace{1.5mm}{ $P_{max}$ in $\%$ and $E_{B-V}$ in mag.} \\
\hspace{3mm}{ o : observed} \\
\hspace{3mm}{ f : foreground} \\
\hspace{3mm}{ i : intracluster}} 
\end{quote}
\end{table}

Figure  4  shows  the  Galactic  distribution {\it(l,b)}  of  all  the
observed stars in the IC 1805.   We have used open circles for all the
observed  stars and filled  circles for  the selected  {\it frontside}
stars. Figures 5a and 5b show the colour excess $E_{B-V}$ plotted as a
function  of  galactic longitude  ${\it(l)}$  and latitude  ${\it(b)}$
respectively, filled  circles for  the selected {\it  frontside} stars
and open circles for the other  members.  

In  order to model  the effect  of foreground  extinction, we  fit the
colour excesses $E_{B-V}$ for the  all nine {\it frontside} stars to a
plane ({\it l,b}). The final equation of the fit is

\begin{equation}
E_{(B-V)}({\it l,b})= 0.0886975{\it l} + 0.0905122{\it b} - 11.2302
(mag.)
\end{equation}
with an rms error of the  unit weight of 0.05.  The variation of the
excesses with  galactic longitude  ${\it(l)}$ is the  most significant
part  of the  Marraco's  model (Marraco  \etal  1993).  The  observed,
modeled and {\it  intracluster} values of the colour  excesses for the
{\it non-frontside} stars are listed  in Table 4.  It can be mentioned
that  the modeled  value of  the frontside  excess led  to  very small
values of the {\it intracluster} excess.

The weighted mean of $\lambda_{max}$ for nine {\it frontside} stars is
\begin{equation}
\overline {\lambda_{max}}= 0.544\pm 0.005 \ \ (\mu m)
\end{equation}
This value is quite similar to Guetter \etal (1989) for the cluster IC
1805 ($0.54 \pm 0.01 \ \mu m $).  Moreover, this value does not differ
more from  the {\it intracluster} value  of $\lambda_{max}$ $(0.541\pm
0.003 \ \mu m)$. These values  are very close to the mean interstellar
value  of  $\lambda_{max}$ 0.56  $\pm$  0.04  $\mu$m (Serkowski  \etal
1975).  Therefore,  we can  conclude that the  characteristic particle
size distribution as  indicated by the polarization study  of stars in
IC 1805 is  essentially the same as that  for the general interstellar
medium.
\begin{figure}
\centering
\includegraphics[scale = .4, trim = 15 15 15 15, clip]{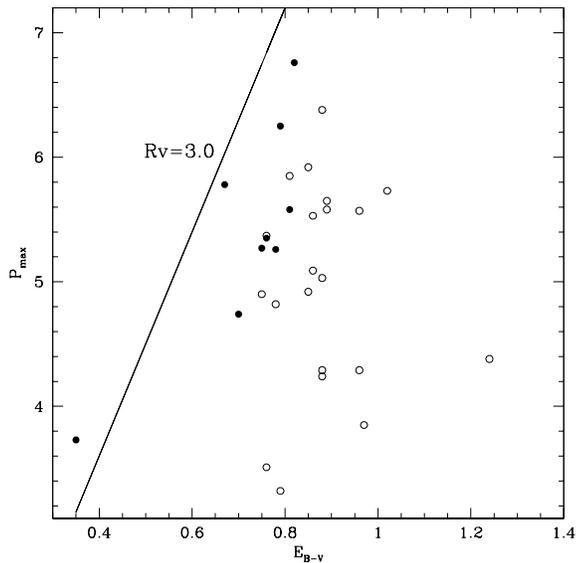}
\vskip.5cm
\caption{Polarization efficiency diagram for the observed dust. Using Rv=3.0, the line of maximum efficiency drawn.
Filled and open circles indicate {\it frontside} and {\it non-frontside} members of IC 1805.}
\label{fig1}
\end{figure}

In  case of polarization,  the weighted  mean of  maximum polarization
$P_{max}$ for nine {\it frontside} stars is
\begin{equation}
\overline {P_{max}}= 4.865\pm 0.022 \ \ (\%)
\end{equation}
and  the  weighted mean  of  maximum  polarization  $P_{max}$ for  the
cluster member is  $5.008\pm 0.005\ (\%)$.  So, there  is also no vast
difference   between   {\it    frontside}   and   {\it   intracluster}
polarization, the  contribution of the {\it  intracluster} material on
polarization  is  $\simeq$ \  $3  \ \%$.   The  weighted  mean of  the
polarization position  angle $\theta_{V}$ for nine  {\it frontside} is
$120.00 \pm  0.06$. It  is possible to  calculate $\overline  {u}$ and
$\overline {q}$,  the mean value  of normalized Stokes  parameters for
the group of nine {\it  frontside} stars.  Subtracting this mean value
from individual $u$ and $q$ for the rest of the observed members of IC
1805 and inverting the procedure,  we obtained the values of $P_{max}$
for {\it intracluster} medium and list them in Table 4.

In diffuse  interstellar medium the polarization  efficiency (ratio of
the  maximum amount  of  polarization to  visual  extinction) can  not
exceed the empirical upper limit,

\begin{equation}
\hspace*{16mm}{P_{max} < 3A_{V} \simeq 3R_{V} \times E_{B-V}}
\end{equation}
which is obtained for interstellar dust particles (Hiltner 1956).  The
ratio  $P_{max}/E_{B-V}$ mainly depends  on the  alignment efficiency,
magnetic  strength  and  also  the  amount of  depolarization  due  to
radiation  traversing more  than one  cloud with  different direction.
Figure  6  shows the  relation  between  $P_{max}$  and colour  excess
$E_{B-V}$  for the  members and  ${\it frontside}$  stars of  IC 1805.
Except the ${\it  frontside}$ star-170, no star lie  at the left block
of the  interstellar maximum line.  Though, the
${\it frontside}$ star-170 falls at the left block of the interstellar
maximum line,  but the  value of $\sigma_1$  for the star-170  is 0.63
(unit  weight  error of  fitting  of  Serkowski polarization  relation
[Eq.(5)]), which is much below than the limiting value 1.6.  So, there
is  very   less  possibility  of  having   intrinsic  polarization  in
star-170. The polarization efficiency plot  for the members of IC 1805
indicates  that apparently  the stars  are not  affected  by intrinsic
polarization. The  dominant mechanism of polarization  in the observed
section  of IC  1805 is  supposed to  be the  alignment of  grains by
magnetic  field,  in  a similar  way  as  that  found in  the  general
interstellar medium.

\section{Conclusions}

The main  results of the  $BVR$ polarimetric study  of IC 1805  can be
summarized as follows:

The  mean value  of maximum  polarization  $P_{max}$ of  IC 1805,  for
foreground  is  $4.865  \pm  0.022  \  \%$ and  while  for  the  ${\it
intracluster}$ medium is  $5.008 \pm 0.005 \ \%$.  The contribution of
the {\it intracluster} dust on polarization is $\simeq$ \ $3 \ \%$.

The mean wavelength dependence of polarization $\lambda_{max}$ for the
cluster  member is  $0.541\pm 0.003  \ \mu  m$ and  for  foreground is
$0.544\pm 0.005 \ \mu m$.  The very less dispersion of $\lambda_{max}$
indicates that the mean {\it intracluster} grain size is quite similar
to general interstellar medium.

Though the observed polarization of some  of the stars in IC 1805 may
be due  to intrinsic stellar  polarization in their  measurements, but
the dominant mechanism is polarization due to the general interstellar
medium.   The  difference between  foreground  and {\it  intracluster}
colour excess  is negligible in the  direction of IC 1805.  There is 
very little  evidence of dust  in the {\it intracluster}  region.  The
highly polarized  and reddened stars may be  projected through various
discrete  foreground clouds, which have different grain sizes and compositions.
	   
\section*{Acknowledgments}

This  research has made  use of  the WEBDA  database, operated  at the
Institute for Astronomy of the University of Vienna, use of image from
the National  Science Foundation and  Digital Sky Survey  (DSS), which
was produced  at the  Space Telescope Science  Institute under  the US
Government grant  NAG W-2166, use  of NASA's Astrophysics  Data System
and   use  of   IRAF,  distributed   by  National   Optical  Astronomy
Observatories, USA. We thank the referee for his constructive comments
which  have lead  to a  considerable  improvement in  the paper.   The
author (BJM) would like to  thanks Orchid, Amitava, Manash, Sanjeev and 
Jessy for their support.

\section*{REFERENCES}
Borgman, J., 1961, Bull. Astron. Inst. Neth., 16, 99 \\
Clayton, Geoffrey C., Cardelli, Jason A., 1988, AJ, 96, 695 \\ 
Coyne, G. V., Magalhaes, A. M., 1979, AJ, 84, 1200 \\
Davis, L. Jr., Greenstein, Jesse L., 1951, ApJ, 114, 206 \\
Guetter, Harry H., Vrba, Frederick J., 1989, AJ, 98, 611 \\
Hiltner, W. A., 1956, ApJS, 2, 389 \\
Ishida, K., 1969, MNRAS, 144, 55 \\	
Johnson, H. L., 1968, Nim.book, 167 \\
Joshi, U. C., Sagar, Ram, 1983, JRASC, 77, 40 \\
Kwon, Suk Minn, Lee, See-Woo, 1983, JKAS, 16, 7 \\
Marraco, H. G., Vega, E. I., Vrba, F. J., 1993, AJ, 105, 258 \\	
Massey, P., Johnson, K. E., Degioia-Eastwood, K., 1995, ApJ, 454, 151 \\
McMillan, R. S., 1978, APJ, 225, 880 \\
Ramaprakash, A. N., Gupta, R., Sen, A. K., Tandon, S. N., 1998, A\&AS, 128, 369\\
Rautela, B. S., Joshi, G. C., Pandey, J. C., 2004, BASI, 32,159 \\	
Sagar, Ram, 1987, MNRAS, 228, 483 \\
Sagar, Ram, Yu, Qian Zhong, 1990, ApJ, 353, 174 \\
Sanders, W. L., 1972, A\&A, 16, 58 \\
Sanders, W. L., 1971, A\&A, 14, 226 \\
Schmidt, G. D., Elston, R., Lupie, O. L., 1992, AJ, 104, 1563 \\
Serkowski, K., 1973, IAUS, 52, 145 \\
Serkowski, K., Mathewson, D. L., Ford, V. L., 1975, ApJ, 196, 261 \\
Shi, H. M., Hu, J. Y., 1999, A\&AS, 136, 313 \\
Tinbergen, J., 1996, Introduction to Astronomical Polarimetry, Cambridge University Press. \\
Turner, D. G., 1976, AJ, 81, 1125 \\
Vasilevskis, S., Sanders, W. L., van Altena, W. F., 1965, AJ, 70, 806 \\
Wilking, B. A., Lebofsky, M. J., Kemp, J. C., Martin, P. G., Rieke, G. H., 1980, ApJ, 235, 905 \\
Whittet, D.C.B., van Breda, I.G., 1978, A\&A, 66, 57 \\

\bsp

\label{lastpage}

\end{document}